\documentclass[%
 aip,
 pop,
 amsmath,amssymb,
 reprint,%
 floatfix
]{revtex4-2}

\usepackage{graphicx}
\usepackage{dcolumn}
\usepackage{bm}
\usepackage[hidelinks]{hyperref}
\usepackage{physics}
\usepackage{esint}
\usepackage[utf8]{inputenc}
\usepackage[T1]{fontenc}
\usepackage{mathptmx}
\usepackage{etoolbox}

\makeatletter
\def\@email#1#2{%
 \endgroup
 \patchcmd{\titleblock@produce}
  {\frontmatter@RRAPformat}
  {\frontmatter@RRAPformat{\produce@RRAP{*#1\href{mailto:#2}{#2}}}\frontmatter@RRAPformat}
  {}{}
}%

\draft
\makeatother

\DeclareMathOperator*{\argmin}{arg\,min}
\newcommand{\qmag}{\mathbf{q}_\mathrm{mag}}

\newcommand{\vv}[1]{\mathbf{#1}}

\begin{document}

\title{
    Hot-spot model for inertial confinement fusion implosions with an applied magnetic field
}

\author{R. Spiers}
\author{A. Bose}
\email{bose@udel.edu}
\author{C. A. Frank}
\affiliation{Department of Physics and Astronomy, University
of Delaware, Newark Delaware 19716, USA}
\author{B. Lahmann}
\author{J. D. Moody}
\author{H. Sio}
\author{D. J. Strozzi}
\affiliation{Lawrence Livermore National Laboratory, 7000 East Ave, Livermore California 94550, USA}

\date{\today}

\begin{abstract}
Imposing a magnetic field on inertial confinement fusion (ICF) implosions magnetizes the electrons in the compressed fuel; this suppresses thermal losses which increases temperature and fusion yield. Indirect-drive experiments at the National Ignition Facility (NIF) with 12 T and 26 T applied magnetic fields demonstrate up to 40\% increase in temperature, 3x increase in fusion yield, and indicate that magnetization alters the radial temperature profile [J.D. Moody \textit{et al.}, Phys. Rev. Lett. \textbf{129}, 195002 (2022), B. Lahmann et al., APS DPP 2022]. In this work, we develop a semi-analytic hot-spot model which accounts for the 2D Braginskii anisotropic heat flow due to an applied axial magnetic field. Firstly, we show that hot-spot magnetization alters the radial temperature profile, increasing the central peakedness which is most pronounced for moderately magnetized implosions (with 8-14 T applied field), compared to both unmagnetized (with no applied field) and highly magnetized (with 26 T or higher applied field) implosions. This model explains the trend in the experimental data which finds a similarly altered temperature profile in the 12 T experiment. Next, we derive the hot-spot model for gas-filled (Symcap) implosions, accounting for the effects of magnetization on the thermal conduction and in changing the radial temperature (and density) profiles. Using this model, we compute predicted central temperature amplification and yield enhancement scaling with the applied magnetic field. The central temperature fits the experimental data accurately, and the discrepancy in the yield suggests a systematic (independent of applied field) degradation such as mix, and additional degradation in the reference unmagnetized shot such as reduced laser drive, increased implosion asymmetry, or the magnetic field suppressing ablator mixing into the hot-spot.
\end{abstract}

\maketitle

\section{Introduction} \label{sec:intro}

In the conventional indirect-drive inertial confinement fusion (ICF) approach, high-energy lasers irradiate a hohlraum, which produces x-rays that ablate the outer surface of a target, imploding a capsule of DT fusion fuel to the thermonuclear conditions required for laboratory fusion gain \cite{atzeniPhysicsInertialFusion2009,nuckollsLaserCompressionMatter1972}. One alternative to conventional ICF imposes an external magnetic field on the capsule to increase the temperature and fusion yield. In a 2011 experiment, an 8 T field was applied at the OMEGA laser facility in the direct-drive configuration, producing a temperature increase of 15\% and a fusion yield improvement of 30\% demonstrating that applied magnetic fields can boost ion temperature and fusion yield \cite{changFusionYieldEnhancement2011a}. An indirect-drive experiment at the National Ignition Facility (NIF) with a 26 T seed field was reported in 2022 to produce a temperature increase of 40\% and a 3x boost in yield \cite{moodyIncreasedIonTemperature2022}. One subsequent indirect-drive experiment applied a modest 12 T applied field and recorded a neutron-averaged temperature increase of 33\% and a 2.5x increase in yield \cite{sioPerformanceScalingApplied2023}. Published in 2024, another NIF experiment using modified laser drive and target characteristics applied a 28 T field and observed a temperature improvement of 28\% and yield increase of 2.19x \cite{strozziDesignModelingIndirectly2024}. These few experiments represent a diverse parameter space of target and laser drive conditions, which illustrates the need to devise one model to accurately predict the temperature and yield amplification in magnetized implosions. In this work, using the standard fluid equations with magnetized thermal conduction, we derive semi-analytic models to explain the trends in amplification of fusion ion temperature and yield due to magnetization.

Each of these magnetized experiments at NIF have been part of the WarmMag platform, which implodes room-temperature gas-filled ``symmetry capsules'' (Symcaps) filled with pure D2 gas and without a dense cryogenic fuel layer. These capsules, which do not come close to igniting and have negligible alpha-heating, are a test platform for studying magnetized thermal conduction.

In these magnetized ICF experiments, the seed B-fields are frozen-in and compressed along with the capsule \cite{gotchevLaserDrivenMagneticFluxCompression2009,sakharovMAGNETOIMPLOSIVEGENERATORS1966}. The magnetic flux is approximately frozen-in to the capsule such that $B_\mathrm{compressed} \sim B_\mathrm{initial} (CR)^2$, where $CR$ is the capsule convergence ratio defined as $R_0 / P_0$. In this expression, $R_0$ is the initial capsule radius and $P_0$ is the hot-spot radius at peak compression, which can be estimated from the zero-order Legendre coefficient of the measured x-ray emission. In the NIF experiments with initial B-fields of 12 T and 26 T, stagnated fields of 6.3 and 7.1 kT are inferred from the secondary DT neutron yield (the 1 MeV T is produced from thermal DD fusion) \cite{sioPerformanceScalingApplied2023,sioDiagnosingPlasmaMagnetization2021}. These strong hot-spot B-fields magnetize the fusion fuel, which suppresses thermal conduction losses and results in a higher hot-spot temperature \cite{walshMagnetizedICFImplosions2022}, boosting the fusion yield \cite{perkinsPotentialImposedMagnetic2017}. Instead of isotropic Spitzer thermal conduction\cite{spitzerTransportPhenomenaCompletely1953}, thermal conductivity in a magnetized plasma is anisotropic (i.e., different heat flux parallel and perpendicular to the B-field), given by Braginskii as \cite{braginskiiTransportProcessesPlasma1965} 
\begin{equation} \label{eq:qanisotropy}
    \mathbf{q}_\mathrm{mag} = -\kappa_\parallel \nabla_\parallel T - \kappa_\perp \nabla_\perp T,
\end{equation}
where $\mathbf{q}_\mathrm{mag}$ is the magnetized heat flux density and $\nabla_\parallel$, $\nabla_\perp$ refer to gradient operations parallel and perpendicular to the direction of the B-field lines. The effect of anisotropy in thermal transport is observed by increased oblateness in magnetized exploding-pusher experiments with strong 50 T applied fields as shown by Bose \textit{et al} \cite{boseEffectStronglyMagnetized2022}. Not included in Eq. \ref{eq:qanisotropy}, in magnetized plasmas the Righi-Leduc effect transports heat perpendicular to the B-field and along isotherms. 2D and 3D simulations indicate this heat flux resulting from hot-spot self-generated magnetic field in conventional ICF implosions (with no applied B-field) can produce $<15 \%$  difference in fusion yield\cite{walshSelfGeneratedMagneticFields2017,walshMagnetizedAblativeRayleighTaylor2022,zhangSelfgeneratedMagneticField2024,frankSelfgeneratedMagneticFields2024}, but this effect is smaller in pre-magnetized implosions since the Righi-Leduc coefficients decrease with high magnetization \cite{walshPerturbationModificationsPremagnetisation2019}.

Reduction in thermal losses is only one of the effects expected from magnetizing ICF implosions. Magnetization also confines fusion products (T and $\alpha$ particles) along the B-field lines which improves alpha-heating localization and may amplify fusion yield in burning-plasma designs \cite{jonesPhysicsBurnMagnetized1986,hoBurnwavePropagationMagnetic2023,libermanDistributionFunctionDiffusion1984,djordjevicIntegratedRadiationMagnetoHydrodynamicSimulations2023}. However, it has been suggested that magnetized thermal conduction suppression may reduce the mass ablation rate of the cold fuel shell, creating a less dense central hot-spot with lower alpha particle stopping power compared to conventional ICF designs \cite{walshPerturbationModificationsPremagnetisation2019}. In addition to thermal conduction suppression and alpha particle stopping, it is speculated that magnetic fields may suppress the amount of degradative high-Z (ablator or fill-tube) mixing \cite{smalyukMeasurementsAblatorGasAtomic2014,hallMeasurementMixFuel2024} into the hot-spot, since the hydrodynamic instabilities contributing to high-Z mix are altered with a B-field \cite{samulskiDecelerationstageRayleighTaylor2022,walshPerturbationModificationsPremagnetisation2019,barbeauDesignHighEnergy2022,sadlerFasterAblativeKelvin2022,walshMagnetizedAblativeRayleighTaylor2022}. Implosions with an especially strong magnetic field have decreased thermal smoothing in the conduction zone and central hot plasma, leading to a less symmetric implosion \cite{boseEffectStronglyMagnetized2022}. These coupled effects of magnetization, including suppression of thermal losses, alteration to alpha particle transport, and change to high-Z mix rate, warrant continued investigation of the magnetized ICF platform to boost fusion gain. 

Previous modeling of magnetized implosions has primarily involved 2-D \cite{strozziDesignModelingIndirectly2024,djordjevicIntegratedRadiationMagnetoHydrodynamicSimulations2023,hoBurnwavePropagationMagnetic2023,perkinsPotentialImposedMagnetic2017,walshMagnetizedICFImplosions2022,oneillBurnPropagationMagnetized2024} and 3-D \cite{walshPerturbationModificationsPremagnetisation2019} radiation-magnetohydrodynamics simulations of specific target designs, and analytic 0-D (where the implosion hot-spot has homogeneous temperature, density, and B-field) modeling of the effect of thermal conduction in a magnetized implosion \cite{walshMagnetizedICFImplosions2022}. While the 2-D and 3-D simulations are most informative, the computational cost means that large scans over target design parameters are less feasible. In contrast, the 0-D model rapidly makes predictions about the temperature and yield increases in magnetized implosions. However, in this work we will show that the simple 0-D physics is insufficient to accurately explain the experimental data. In this work, we develop an analytic model for the effect of magnetized thermal conduction in a magnetized hot-spot, which aims to capture the essential part of the 2-D magnetized implosion physics without the computational cost. These types of reduced-order (spherically symmetric 1-D) models are useful because they can be compared to experimental diagnostics measurements \cite{cerjanIntegratedDiagnosticAnalysis2013a,zylstraBurningPlasmaAchieved2022} or used to analytically describe hot-spot physics processes, such as implosion dynamics \cite{bettiHotspotDynamicsDecelerationphase2001,bettiDecelerationPhaseInertial2002}, ignition characteristics \cite{kishonyInertialConfinementFusion1997,bettiThermonuclearIgnitionInertial2010,daughtonInfluenceMassAblation2023}, the onset of thermonuclear burn waves \cite{christophersonTheoryIgnitionBurn2020a} and mixing processes \cite{meaneySeparatedReactantMix2024}. 

A long-standing analytic model for the hot-spot of conventional unmagnetized ICF implosions involves solving a simplified version of the temperature equation in the steady-state (i.e., $dT/dt = 0$) which results in the profile
\begin{equation} \label{eq:27profile}
    T(\hat{r}) = T_0 (1 - \hat{r}^2)^{2/7},
\end{equation}
where $T_0$ is the temperature at the center of the hot-spot, and $\hat{r} = r / R_\mathrm{hs}$ where $R_\mathrm{hs}$ is the radius of the hot-spot \cite{lindlInertialConfinementFusion1998}. This derivation, with generalization to magnetized implosions, is presented in Sec. \ref{sec:lindl}.

\begin{figure}[h]
    \centering
    \includegraphics{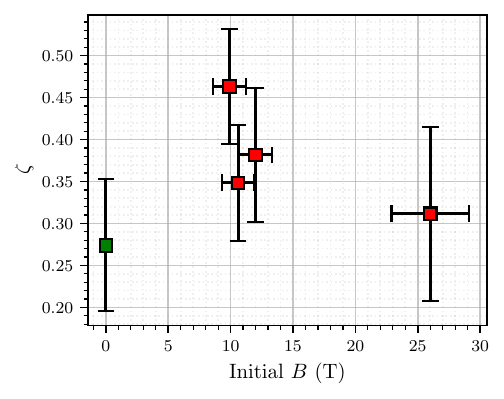}
    \caption{Inferred temperature exponent $\zeta$ based on the temperature profile $T = T_0 \left( 1 - (r/R_\mathrm{hs})^2 \right)^\zeta$ (Eq. \ref{eq:zetaprofile}) for three indirect-drive experiments at the National Ignition Facility with room-temperature $\mathrm{D_2}$ targets. The data used arises from shots N210607, N210912, and N220110 \cite{sioPerformanceScalingApplied2023} and the plot is reproduced with data from reference \onlinecite{lahmannOneDimensionalPlasma2022}.}
    \label{fig:brandonplot_noprediction}
\end{figure}

Recent analysis of NIF experiments with an applied B-field indicates that magnetized hot-spots cannot be accurately fit by the profile in Eq. \ref{eq:27profile}. The measured quantities, including DD neutron yield $Y_{DDn}$, neutron-averaged ion temperature $\langle T_i \rangle_{DDn}$, X-ray penumbral imaged radii in the plane with the applied field ($P_0$) or perpendicular to it ($M_0$), burn width, and total X-ray emission, were input to a 1D hot-spot model to infer the central ion temperature $T_0$, the hot-spot radius $R_\mathrm{hs}$, and the hot-spot temperature profile shape \cite{lahmannOneDimensionalPlasma2022}. The shape of the radial temperature profile was parameterized by the variable $\zeta$ in Eq. \eqref{eq:zetaprofile}.
\begin{equation} \label{eq:zetaprofile}
    T(\hat{r}) = T_0 (1 - \hat{r}^2)^\zeta
\end{equation}
The resulting $\zeta$ inferred from NIF shots N210607 (26 T), N210912 (0 T), and N220110 (12 T, penumbral X-ray measurements from three different lines of sight) is shown in Fig. \ref{fig:brandonplot_noprediction} with error-bars arising from uncertainties associated with the measured quantities (e.g., ion temperature and burn width).

Fig. \ref{fig:brandonplot_noprediction} indicates that the temperature profile shape is altered for magnetized implosions. However, experimental uncertainty means that conclusions about the profile shape, such as whether $\zeta$ increases monotonically with the applied B-field, cannot be determined by these experiments alone. In this work, we first develop a \textit{simple} and \textit{advanced} model for the magnetized hot-spot temperature profile, i.e., $T(r)$. We call these models ``1.5-dimensional'', because their hydrodynamic variables (density, temperature) are 1D (spherically symmetric) but the magnetized heat flux (Eq. \ref{eq:qanisotropy}) is treated in 2D---parallel and perpendicular to the magnetic field. After obtaining the hot-spot profiles $T(r)$ as a function of hot-spot magnetization, we couple these to an evolutionary hot-spot model for the magnetized NIF Symcaps (without a cryogenic dense fuel layer) which self-consistently evolves the dynamic hot-spot quantities (Sec. \ref{sec:Tamp}). The results of this model are compared against experimental measurements of the central temperature and yield and the implications for agreement and discrepancy are discussed.

\section{Simple model} \label{sec:lindl}

One method to derive the temperature profile includes looking for self-similar solutions to the electron temperature ($T_e$) equation. 
\begin{equation} \label{eq:tempequation}
    \frac{3}{2} n_e \frac{\partial T_e}{\partial t} = - \nabla \cdot \qmag + P_\alpha - P_\mathrm{rad} + P_W
\end{equation}
This equation describes the time-evolution of the electron temperature due to magnetized thermal conduction and the power densities associated with alpha heating $P_\alpha$, Bremsstrahlung cooling $P_\mathrm{rad}$, and compressive work $P_W$. In this work, we neglect the contribution of ion thermal conduction since it is much less than electron conduction unless $Z$ (ion charge) is high or a strong B-field is applied, neither of which is the case for current magnetized-ICF experiments. For the remainder of this work, we assume $T_i = T_e \equiv T$ for simplicity in the models.

There is a closed-form solution to Eq. \ref{eq:tempequation} for an unmagnetized hot-spot temperature profile, given by Ref. \onlinecite{lindlInertialConfinementFusion1998}. This derivation involves separation of variables (into spatial and temporal parts), and that the total power density $P_\alpha - P_\mathrm{rad} + P_W \equiv P_\mathrm{tot}$ is spatially homogeneous throughout the hot-spot. Using these same assumptions, we can derive an equation for the temperature profile in a magnetized hot-spot. First, we integrate the temperature equation over a subsphere within the hot-spot with radius $r$.
\begin{equation} \label{eq:lindlintegral}
    4 \pi \int_0^{r} {r^\prime}^2 P_\mathrm{tot} dr^\prime = \oiint_{r^\prime=r} \mathbf{q}_\mathrm{mag} \cdot d\mathbf{S}^\prime
\end{equation}
Since the heating power density is assumed homogeneous, the left-hand-side integral evaluates to $(4 \pi r^3 / 3) P_\mathrm{tot}$. With no magnetic field, the right side evaluates to $-4 \pi r^2 \kappa_0 T^{5/2} (\partial T / \partial r)$, using the usual Spitzer form of the thermal conductivity \cite{spitzerTransportPhenomenaCompletely1953}. Equating these terms and applying the boundary conditions $T(r = 0) = T_0$ and $T(r = R_\mathrm{hs}) = 0$ gives an ordinary differential equation with closed-form solution given by Eq. \ref{eq:27profile}. 

For implosions with an applied B-field, magnetization of the fusion fuel suppresses the heat flow $\qmag$, since the electrons are confined to their gyroradius, inhibiting trajectories perpendicular to B-field lines. Neglecting the Righi-Leduc effect since the hydrodynamic variables are assumed to be axisymmetric, the magnetized heat flux $\qmag$ using the Braginskii coefficients \cite{braginskiiTransportProcessesPlasma1965} for a $Z=1$ plasma is given to be:
\begin{align} \label{eq:bragheat}
    \qmag &= -\kappa_0 T^{5/2} \left( \nabla_\parallel T + \frac{\kappa_\perp}{\kappa_{\parallel}} \nabla_\perp T \right) \\
    \frac{\kappa_\perp}{\kappa_{||}} &= \frac{1.475 \chi^2 + 3.770}{\chi^4 + 14.79 \chi^2 + 3.770}
\end{align}
For singly-ionized ($\bar{Z} = 1$) plasmas, $\kappa_0 = 9.81 \cdot 10^{19}/\lambda_{ei} \ \mathrm{erg/(cm\ s\ keV^{7/2})}$, although fits for arbitrary $\bar{Z}$ are available in Refs. \onlinecite{sadlerSymmetricSetTransport2021,epperleinPlasmaTransportCoefficients1986}. The thermal suppression factor $\kappa_\perp/\kappa_\parallel$ depends on the electron Hall parameter $\chi \equiv \omega_e \tau_{ei}$. Expanding the electron gyrofrequency $\omega_e$ and the electron-ion collision time $\tau_{ei}$ in the ICF-relevant regime for temperatures and densities \cite{richardson2019NRLPlasma2019} yields
\begin{equation} \label{eq:hallparameter}
    \chi = 6.05 \cdot 10^{16} \frac{T_e^{3/2} B}{n_e \lambda_{ei}},
\end{equation}
where $T_e$ is the electron temperature in eV, $B$ is the magnetic field in T, $n_e$ is the electron number density in $\mathrm{cm}^{-3}$, and $\lambda_{ei}$ is the Coulomb logarithm, which in this regime can be approximated as \cite{richardson2019NRLPlasma2019}: $\lambda_{ei} \approx 24 - \ln{\left( n_e^{1/2} T_e^{-1} \right)}$.

In Eq. \ref{eq:bragheat}, $\nabla_\parallel$ and $\nabla_\perp$ denote spatial differentiation parallel and perpendicular to the B-field, respectively. Mathematically, this means that for a B-field in the cylindrical $z$ direction (the unit vector $\mathbf{e}_z$), then $\nabla_\parallel T \equiv \left( \mathbf{e}_z \cdot \nabla T \right) \mathbf{e}_z$ and $\nabla_\perp T \equiv \mathbf{e}_z \times \left( \nabla T \times \mathbf{e}_z \right)$ are the spatial gradient operations \cite{walshMagnetizedICFImplosions2022}. In order to substitute the magnetized $\mathbf{q}_\mathrm{mag}$ (Eq. \ref{eq:bragheat}) into the temperature profile model (Eq. \ref{eq:lindlintegral}), needed is $\mathbf{q}_\mathrm{mag} \cdot d\mathbf{S} = (\mathbf{q}_\mathrm{mag} \cdot \mathbf{e}_r) r^2 \sin{\theta} d\theta d\phi$. 

To make the integration over $d\vv{S}$ analytically tractable, we consider only the zeroth-order expansion of the temperature profile over the polar angle, i.e., $T(r, \theta) = \bar{T}(r) + \delta T(r, \theta) \approx \bar{T}(r)$. We note that this radial temperature profile $\bar{T}(r)$ is not, in general, a polar-angle average of the 2D temperature profile $T(r, \theta)$, nor is $\bar{T}(r)$ in itself a steady-state solution to Eq. \ref{eq:tempequation} when $B \neq 0$. However, since $\bar{T}(r)$ is bounded ($T(r, \frac{\pi}{2}) \leq \bar{T}(r) \leq T(r, 0)$), we can consider it a reasonable estimation of the radial temperature profile and it lies somewhere between the polar lineout and the equatorial lineout. For notational convenience, throughout the rest of this paper the variable $T$ implies $\bar{T}(r)$ unless explicitly stated otherwise. Pursuant to this expansion, then the gradient of $T$ is given by $\nabla T = (\partial T/ \partial r) \mathbf{e}_r$ and the expressions for the spatial derivatives in the radial direction follow.
\begin{subequations}
    \begin{equation} \label{eq:nablapar}
        \left( \nabla_{||} T \right) \cdot \mathbf{e}_r = \left( \left(\mathbf{e}_z \cdot \frac{\partial T}{\partial r} \mathbf{e}_r \right) \mathbf{e}_z \right) \cdot \mathbf{e}_r = \frac{\partial T}{\partial r} \cos^2 \theta
    \end{equation}
    \begin{equation}  \label{eq:nablaperp}
        \begin{split}
            \left( \nabla_\perp T \right) &\cdot \mathbf{e}_r \\
            &= \left( \mathbf{e}_z \times \left( \frac{\partial T}{\partial r} \mathbf{e}_r \times \mathbf{e}_z \right) \right) \cdot \mathbf{e}_r \\
            &= \frac{\partial T}{\partial r} \left( \mathbf{e}_r - \left( \mathbf{e}_r \cdot \mathbf{e}_z \right) \mathbf{e}_z \right) \cdot \mathbf{e}_r = \frac{\partial T}{\partial r} \sin^2 \theta
        \end{split}
    \end{equation}
\end{subequations}
In these equations, $\theta$ is the spherical polar angle between $\mathbf{e}_z$ and $\mathbf{e}_r$ which projects the temperature gradient onto the axes parallel or perpendicular to the $B$-field. The total heat flow in the radial direction is then calculated using Eq. \ref{eq:bragheat}.
\begin{equation} \label{eq:qangdep}
    \mathbf{q}_\mathrm{mag} \cdot \mathbf{e}_r = - \kappa_0 T^{5/2} \frac{\partial T}{\partial r} \left( \cos^2 \theta + \frac{\kappa_\perp}{\kappa_{||}} \sin^2 \theta \right)
\end{equation}
The expression from \eqref{eq:qangdep} is inserted into the integral on the right-hand side of Eq. \ref{eq:lindlintegral}, where the integration is carried out assuming the scalar quantity $\kappa_\perp/\kappa_\parallel$ (which is a function of the Hall parameter $\chi$ only) is spherically symmetric (i.e., that the profile of $|\vv{B}|$ is a function of $r$ only).
\begin{equation} \label{eq:unnormalizedlindl}
    \frac{1}{3} r P_\mathrm{tot} = -\kappa_0 T^{5/2} \frac{\partial T}{\partial r} \left( \frac{1}{3} + \frac{2}{3} \frac{\kappa_\perp}{\kappa_\parallel} \right).
\end{equation}
This differential equation must be numerically integrated since $\kappa_\perp/\kappa_\parallel$ depends on $\chi$ which is a function of the hot-spot profiles $T(r)$ and $B(r)$. By assuming that the hot-spot is isobaric with an ideal gas equation of state, and neglecting magnetic pressure since it is much less than the thermal pressure in this regime, then $n \sim T^{-1}$ and therefore the profile of the Hall parameter scales as $\chi(r) \sim [T(r)]^{5/2} B(r)$, where we have neglected the contribution from the Coulomb logarithm for simplicity.

Usage of a numerical solver on equation \eqref{eq:unnormalizedlindl} requires normalization of the variables. We will show that solving Eq. \eqref{eq:unnormalizedlindl} in normalized variables gives a unique solution to the radial temperature profile shape for each value of the central Hall parameter $\chi_0$. Let $T = T_0 \hat{T}$, $r = R_{hs} \hat{r}$, $\chi = \chi_0 \hat{T}^{5/2} \hat{B}$, where $B = B_0 \hat{B}$ such that $\hat{B}(\hat{r} = 0) = 1$. The boundary conditions $T(r=0) = T_0$ and $T(r=R_{hs}) = 0$ then become $\hat{T}(\hat{r} = 0) = 1$ and $\hat{T}(\hat{r} = 1) = 0$. This second boundary condition can only be satisfied with a total power density $P_\mathrm{tot}$ which exactly balances the thermal conduction out of the hot-spot to maintain a steady-state temperature. Mathematically, this means we impose the normalization $P_\mathrm{tot} = 3A \kappa_0 T_0^{7/2}/R_\mathrm{hs}^2$, where $A$ is a dimensionless parameter which depends on $\chi_0$. For the unmagnetized ($\chi_0 = 0$) case, it can be shown with pen and paper that $A = 4/7$ is required for a steady-state temperature profile \cite{kishonyInertialConfinementFusion1997}, where one would separate variables in the first line of Eq. \ref{eq:normlindl}, integrate from the hot-spot center to an arbitrary radius, and satisfy the boundary condition $\hat{T}(1)=0$. Implementing the above normalizations into Eq. \eqref{eq:unnormalizedlindl} along with the expansion for $\kappa_\perp/\kappa_\parallel$ gives Eq. \eqref{eq:normlindl}.
\begin{multline} \label{eq:normlindl}
    A \hat{r} = - \hat{T}^{5/2} \frac{\partial \hat{T}}{\partial \hat{r}} \\
    \times \left( \frac{1}{3} + \frac{2}{3} \frac{1.475 \chi_0^2 \hat{T}^5 \hat{B}^2 + 3.770}{\chi_0^4 \hat{T}^{10} \hat{B}^4 + 14.79 \chi_0^2 \hat{T}^5 \hat{B}^2 + 3.770}\right)
\end{multline}
Subject to the two-point boundary values we have imposed on the system, Eq. \ref{eq:normlindl} can be solved using the Shooting method by viewing the system as an eigenvalue equation for eigenvalues $A$. More specifically, we use a fourth-order Runge-Kutta solver initialized with $\hat{T}(0) = 1$ and use a root-finding algorithm to identify $A$ such that the radius when the temperature reaches zero is the hot-spot radius (i.e., $\hat{T}(1) = 0$). When we use this solver assuming a spatially homogeneous B-field profile, we find that the value of $A$ decreases as $\chi_0$ increases, meaning less power is required to maintain steady-state temperature due to magnetic insulation of the hot-spot. A simple fit to the numerical solution gives 
\begin{equation}
    A \approx \frac{4}{21} \left( 1 + \frac{2}{1 + 0.89 \chi_0 + 0.059 \chi_0^2} \right),
\end{equation}
demonstrating that the required heating to maintain a steady-state temperature in a highly magnetized hot-spot is $1/3$ the amount needed for an unmagnetized hot-spot. In addition, $A$ depends on the choice of B-field profile $\hat{B}$, which simulations show is typically not spatially homogeneous but rather is strongest in the hot-spot center and decreases towards the shell \cite{changFusionYieldEnhancement2011a,sioPerformanceScalingApplied2023,strozziDesignModelingIndirectly2024}. These non-uniform B-fields are discussed at the end of this section.

\begin{figure}[h]
    \centering
    \includegraphics{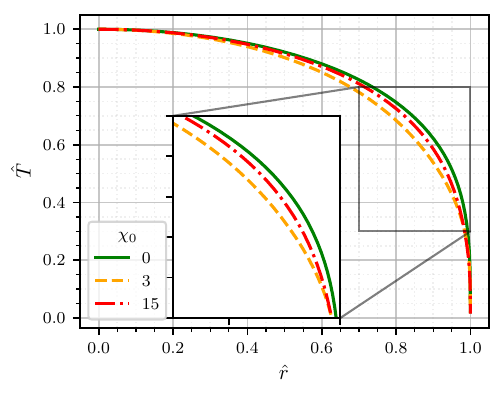}
    \caption{Normalized temperature profiles corresponding to the solutions of Eq. \ref{eq:normlindl} for $\chi_0 = 0$ (solid green line), $\chi_0 = 3$ (dashed yellow curve), and $\chi_0 = 15$ (dot-dashed red curve) for the spatially homogeneous B-field profile $\hat{B} = 1$.}
    \label{fig:Tvsrhatdifferentchi0}
\end{figure}

\begin{figure}[h]
    \centering
    \includegraphics{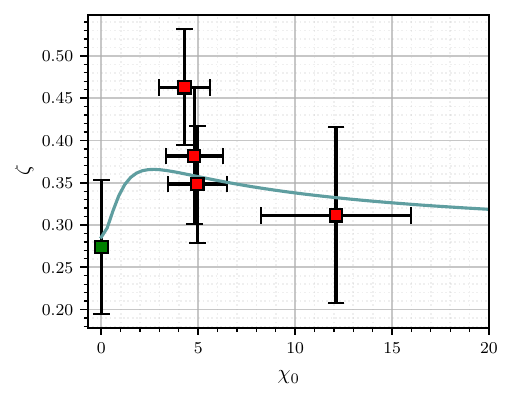}
    \caption{Temperature profile exponent $\zeta$ corresponding to a solution of equation \eqref{eq:normlindl} with homogeneous B-field profile $\hat{B} = 1$. 
    The experimental data arises from Refs. \onlinecite{sioPerformanceScalingApplied2023,lahmannOneDimensionalPlasma2022} with propagated error-bars corresponding to uncertainties in the measured values.}
    \label{fig:zetaoverlayhomoB}
\end{figure}

Shown in Fig. \ref{fig:Tvsrhatdifferentchi0} is that the solution corresponding to the moderately magnetized $\chi_0 = 3$ case is more peaked (i.e., has a temperature profile which begins to decrease at smaller $\hat{r}$) than the unmagnetized solution. Once the temperature profile is solved for a given $\chi_0$, we perform a non-linear curve fit to find the temperature profile exponent $\zeta$ (from Eq. \ref{eq:zetaprofile}) for that profile. The solutions $\zeta(\chi_0)$ are shown in Fig. \ref{fig:zetaoverlayhomoB}, which aligns with the NIF experimental measurements and explains why the temperature profile exponent first increases to a maximum at $\chi_0 \approx 3$ before decreasing.

The results of this model are shown in Fig. \ref{fig:zetaoverlayhomoB} with the inferred temperature profile exponent $\zeta$ and central magnetization $\chi_0$ in the NIF experiments and corresponding error-bars. Uncertainty in the experimental values stems from measurement uncertainty, mainly associated with the duration of the burn width and the compressed magnetic field \cite{lahmannOneDimensionalPlasma2022}.

In Fig. \ref{fig:zetaoverlayhomoB}, the analytic theory and experiments both exhibit an increase in the temperature profile exponent $\zeta$ for the moderately magnetized implosions. However, there is one line-of-sight measurement of shot N220110 with $\zeta = 0.46$ which is not quantitatively fit by the analytic model. Demonstrated in Sec. \ref{sec:betti} is that by including additional physics effects into the analytic model, this measurement may be more accurately fit.

Solutions of the hot-spot profile shape were also carried out assuming other profiles for the B-field that were not spatially homogeneous in magnitude but still uni-directional (i.e., $\mathbf{B} = B(r) \vec{e}_z$). Results from both the 1D magnetohydrodynamics code LILAC-MHD \cite{changFusionYieldEnhancement2011a} and recent 2-D simulations using the rad-hydro MHD code \texttt{Lasnex} indicate that the B-field is peaked at the center of the hot-spot and decreases to some fraction of its maximum at the hot-spot boundary; \texttt{Lasnex} simulations also show that the field lines become twisted \cite{sioPerformanceScalingApplied2023,strozziDesignModelingIndirectly2024}. Depending on how the field lines are twisted during the implosion, the insulating effect of the magnetized thermal conduction is altered and may produce a different temperature profile. The change in temperature profile from the non-axial B-field configuration would alter the integration on the right-hand-side of Eq. \ref{eq:lindlintegral}, meaning that Eq. \ref{eq:unnormalizedlindl} would have the coefficients in the parentheses $\frac{1}{3}$ and $\frac{2}{3}$ be adjusted according to the direction of the B-field as a function of the polar angle $\theta$. In addition, this simple model can study the effects of a spatial inhomogeneity in the B-field. Using the approximate B-field profile suggested by the LILAC-MHD and \texttt{Lasnex} simulations ($\mathbf{B} = B_0(1 - 0.5\hat{r})\vv{e}_z$) to solve Eq. \ref{eq:normlindl} produces similar results to Fig. \ref{fig:zetaoverlayhomoB} except the horizontal axis should be replaced by $\langle \chi \rangle$. Here, $\langle \chi \rangle$ designates the volume-average Hall parameter instead of the central Hall parameter, since for the same central B-field and temperature, this change to the B-field profile reduces the volume-average Hall parameter by 37\%. Since the qualitative behavior of the temperature profile is shown here to be independent of the choice of B-field profile, we use the homogeneous B-field profile in the following sections to make the equations tractable.

\section{Advanced model} \label{sec:betti}

The simple model presented in the previous section simplifies the underlying physics to obtain a solution to the temperature profile that is nearly analytic (aside from the final integration). In contrast, the advanced model in this section includes the effects of mass ablation into the hot-spot and solves the hydrodynamic equations in Lagrangian form, leading to an equation which describes the shell dynamics and does not require the assumption of a steady-state hot-spot (i.e., $dT/dt \neq 0$). Specifically, we introduce magnetized thermal conduction effects into the hot-spot model from Betti \textit{et al.} \cite{bettiHotspotDynamicsDecelerationphase2001}. Since this model includes mass ablation, its results are more applicable to future implosions with a cryo-layered DT shell rather than the room-temperature Symcaps used in magnetized NIF experiments to-date. We will, however, show that this model may still fit the Symcap data better than the simple model, owing to the fact that our process focuses on the center of the hot-spot, where the effects of mass ablation are expected to be small.

Just as with the previous simple model, the temperature profile is 1-dimensional while the heat flux $\qmag$ is treated in 2D. However, due to the complexity of the underlying hot-spot model we generate a fit to the magnetized heat flux using a Spitzer-type power law dependent on the magnetization.
\begin{equation} \label{eq:heatapprox}
    \mathbf{q}_\mathrm{apx} = -\kappa_\mathrm{eff} T^\nu \nabla T
\end{equation}
For standard (unmagnetized) Spitzer electron conduction \cite{spitzerTransportPhenomenaCompletely1953} this results in $\kappa_\mathrm{eff} = \kappa_0$ and $\nu = 5/2$. Treating exotic (non-Spitzer) thermal conduction by adjusting the coefficient $\kappa_\mathrm{eff}$ and the exponent $\nu$ according to the plasma conditions has precedent in modeling both magnetized \cite{walshMagnetizedICFImplosions2022} and unmagnetized \cite{patelHotspotConditionsAchieved2020} implosions.

We fit $\kappa_\mathrm{eff}$ and $\nu$ in Eq. \ref{eq:heatapprox} by equating the 
exact magnetized heat flux $\mathbf{q}_\mathrm{mag}$ and the approximated power-law heat flux $\mathbf{q}_\mathrm{apx}$ through a sphere.
\begin{equation} \label{eq:matchheatflux}
    \oiint_{r=r^\prime} \mathbf{q}_\mathrm{mag} \cdot d\mathbf{S} \approx \oiint_{r=r^\prime} \mathbf{q}_\mathrm{apx} \cdot d\mathbf{S}
\end{equation}
Eq. \ref{eq:matchheatflux} represents a matching condition where the total heat flux through a sphere of radius $r < R_\mathrm{hs}$ is equivalent whether one uses the exact anisotropic heat flow equation or the approximated isotropic heat flow equation with appropriate choice of $\kappa_\mathrm{eff}$ and $\nu$. However, it is impossible to exactly match the heat flux for all radii $r$ with only the two variables $\nu$ and $\kappa_\mathrm{eff}$. Instead, we use least-squares minimization to determine $\nu$ and $\kappa_\mathrm{eff}$ which accrues the least error for all $r \in (0, R_\mathrm{hs}]$. Formally, the least-squares minimization of a function $f(x)$ which has parameters $a$ and $b$ (e.g., $f(x) = a + bx - 3$) would be written as $\argmin_{(a, b)} \sum_x \left( f(x) \right)^2$ (e.g., $\argmin_{a, b} \sum_x \left( a+bx-3 \right)^2 \Rightarrow (a,b)=(3,0)$). Therefore, we can write the matching condition for Eq. \ref{eq:matchheatflux}, where we find $\kappa_\mathrm{eff}$ and $\nu$ such that Eq. \ref{eq:matchheatflux} is approximately satisfied for all $r$, in the context of least-squares minimization, as:
\begin{equation} \label{eq:argmingeneral}
    \argmin_{(\kappa_\mathrm{eff}, \nu)} \int_0^{R_\mathrm{hs}} \left( \left[ \oiint \left( \mathbf{q}_\mathrm{mag} - \mathbf{q}_\mathrm{apx} \right) \cdot \frac{d\mathbf{S}}{{r}^2} \right]_{r=r^\prime}  \right)^2 dr^\prime.
\end{equation}
This expression finds the optimal $\kappa_\mathrm{eff}$ and $\nu$ which minimize the square of the residuals (i.e., the difference between the actual heat flux through a sphere of radius $r^\prime$ and the approximated heat flux through the same sphere). Instead of minimizing the difference of the actual heat fluxes, which would give preferential weight to the larger spheres, we normalize the flux to the sphere surface area (i.e., $d\mathbf{\hat{S}} = d\mathbf{S}/r^2$) to obtain a better fit to the overall data. We have already computed this surface integral of $\mathbf{q}_\mathrm{mag}$ in Eq. \ref{eq:unnormalizedlindl}, and a simple calculation yields that the corresponding surface integral of $\mathbf{q}_\mathrm{apx}$ is $-4 \pi (r^\prime)^2 \kappa_\mathrm{eff} T^\nu (\partial T/\partial r)$. These integrals are inserted into Eq. \ref{eq:argmingeneral} along with the variable normalizations $\hat{T} = T/T_0$, $\hat{r} = r/R_\mathrm{hs}$, and $\hat{\kappa}_\mathrm{eff} = \kappa_\mathrm{eff} T_0^{\nu-2.5} / \kappa_0$ to arrive at Eq. \ref{eq:argminnormalized}.
\begin{multline} \label{eq:argminnormalized}
    \argmin_{(\kappa_\mathrm{eff}, \nu)} \int_0^1 \left( \left[ \hat{T}^{5/2} \frac{\partial \hat{T}}{\partial \hat{r}} \left( \frac{1}{3} + \frac{2}{3} \frac{\kappa_\perp}{\kappa_\parallel} \right) \right. \right.\\
     \left. \left.- \hat{\kappa}_\mathrm{eff} \hat{T}^\nu \frac{\partial \hat{T}}{\partial \hat{r}} \right]_{\hat{r} = r^\prime} \right)^2 dr^\prime
\end{multline}
Eq. \ref{eq:argminnormalized} is dimensionless and independent of the central temperature $T_0$. The first term within the brackets is related to $\mathbf{q}_\mathrm{mag}$ and has functional dependencies on $\zeta$ and $\chi_0$, while the second term is related to $\mathbf{q}_\mathrm{apx}$ and is dependent on $\zeta$, $\kappa_\mathrm{eff}$, and $\nu$. To carry out this least-squares minimization, a few assumptions must be made and equations solved.
\begin{enumerate}
    \item Assume a form for the temperature profile: \\$\hat{T}(r) = (1-\hat{r}^2)^\zeta$.
    \item Using a hot-spot physics model, find the relationship between $\zeta$ and $\nu$.
    \item Use a least-squares minimizer to find the optimal $\kappa_\mathrm{eff}$ and $\nu$ for each $\chi_0$.
    \item Given $\nu(\chi_0)$, calculate $\zeta(\chi_0)$.
\end{enumerate}

In the simple model section, we found that $\hat{T}(r) = (1-\hat{r}^2)^\zeta$ fit the data in general, so we choose it as our ansatz for this section. Step (2) encodes the physics processes, including magnetized thermal conduction, alpha heating and $pdV$ work, and mass ablation. The rest of this section is laid out as follows. First, we validate the four-step least-squares method by solving the simple model over again with the four-step method. Then, we perform Step (2) for the advanced model which includes mass ablation and the non-steady-state temperature equation. Finally, we calculate Steps (3) and (4) and discuss the results.

To prove that this four-step method works, we solve the simple model (Sec. \ref{sec:lindl}) again, but instead of numerically integrating the temperature equation we instead use the four-step least-squares method. First, we replace $\mathbf{q}_\mathrm{mag}$ with $\mathbf{q}_\mathrm{apx}$ in Eq. \eqref{eq:tempequation}. Using the same assumptions and normalizations as in Sec. \ref{sec:lindl}, it can be shown that, instead of Eq. \ref{eq:normlindl}, the resulting equation is given by
\begin{equation} \label{eq:lindlthebettiway}
    A \hat{r} = -\hat{\kappa}_\mathrm{eff} \hat{T}^\nu \frac{\partial \hat{T}}{\partial \hat{r}}.
\end{equation}
This equation has an analytic solution arising from separation of variables, given by $A = \hat{\kappa}_\mathrm{eff}$ and $\hat{T} = (1 - \hat{r}^2)^{1/(1+\nu)}$, from which we can identify that $\zeta = 1/(1+\nu)$. Now that we know the relationship between $\zeta$ and $\nu$ for this hot-spot model, we can solve Steps (3) and (4) and find that the two methods of producing $\zeta(\chi_0)$ agree within 1\% error, illustrating that both methods produce the same solution given this choice of temperature ansatz.

Now we move on to finding the relationship between $\zeta$ and $\nu$ for the advanced Betti \textit{et al.} hot-spot model \cite{bettiHotspotDynamicsDecelerationphase2001} when including magnetized thermal conduction. In their work, the temperature profile is found by numerically integrating a differential equation of dimensionless variables $F$ and $\varphi$ which are related to the physical parameters $\hat{T}$ and $\hat{r}$.
\begin{equation} \label{eq:bettiODE}
    \frac{1}{\nu + 1} \varphi + F^{4/3} \left( \frac{dF}{d\varphi} \right)^{\nu-2} \frac{d^2 F}{d\varphi^2} = 0
\end{equation}
The hot-spot temperature profile is related to the solution $F(\varphi)$ by the following relations: $\frac{dF}{d\varphi} (\varphi_0) = 0$, $\hat{r}^{3} = F(\varphi)/F(\varphi_0)$, and $\hat{T} = (dF/d\varphi)$. With these equations, the hot-spot profile $\hat{T}(\hat{r})$ can be found provided any $\nu$. To find the resulting $\zeta$, we generalize the technique that was used in Ref. \onlinecite{bettiHotspotDynamicsDecelerationphase2001} for $\nu = 5/2$. In their work, they find that the solution to the differential equation in Eq. \ref{eq:bettiODE} (the temperature profile) is similar to the standard $(1-\hat{r}^2)^{2/7}$ profile for $0<\hat{r}<0.8$ but is altered near the hot-spot edge due to mass ablation, so they modify the equation to the form,
\begin{equation} \label{eq:bettidelta}
    \hat{T} = \frac{(1 - \hat{r}^2)^{(1/\nu)}}{1 - \delta r^2},
\end{equation}
where $\delta$ is a free parameter to be fit. For the unmagnetized case, they found that $\delta \approx 0.15$ fits the solution with the classical Spitzer $\nu = 5/2$. This form of the temperature profile was chosen since it provides a smooth transition between taking the shape of the simple model $\hat{T} \approx (1-\hat{r}^2)^{2/7}$ for $0 < \hat{r} < 0.8$, but near the outer part of the hot-spot the slope is less steep due to mass ablation, where the profile takes a shape similar to $\hat{T} \sim (1-\hat{r}^2)^{2/5}$.

\begin{figure}[h]
    \centering
    \includegraphics{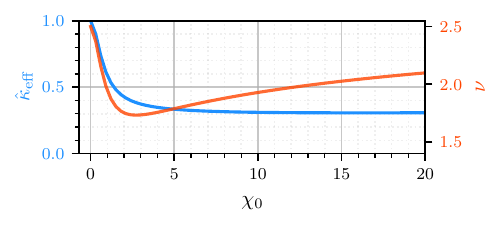}
    \caption{Solutions to the minimization expression in Eq. \ref{eq:argminnormalized} for $\zeta(\nu)$ derived using the advanced model. $\hat{\kappa}_\mathrm{eff}$ is shown on the left axis colored solid blue and $\nu$ is the dotted orange curve on the right axis.}
    \label{fig:keffnu}
\end{figure}

The four-step process we describe could easily be extended to study the outer region of the hot-spot, where the mass ablation effects are important, however in this paper we are instead interested in characterizing the room-temperature magnetized experiments without a cryogenic fuel layer that undergoes mass ablation. For the central region of the hot-spot ($0 < \hat{r} < 0.8$), the primary difference between the simple and advanced models is that the temperature is not steady-state ($\partial T/\partial t \neq 0$) which makes this model more applicable. Thus, by choosing in this analysis to neglect the part of the hot-spot most cooled by mass ablation, we re-fit the main section of the hot-spot ($0 < \hat{r} < 0.8$) to the profile $\hat{T} = (1 - \hat{r}^2)^\zeta$. Overall, this methodology allows us to find $\zeta$ as a function of $\nu$ for the advanced hot-spot model with additional physics effects included.
\begin{equation} \label{eq:bettizetavsnu}
    \zeta \approx \frac{0.86}{0.49 + \nu}
\end{equation}
This concludes Step (2) for the advanced model. In accordance with Step (3), Eq. \ref{eq:bettizetavsnu} is inserted into the minimization expression (Eq. \ref{eq:argminnormalized}) to find $\kappa_\mathrm{eff} (\chi_0)$ and $\nu(\chi_0)$, the results of which are shown in Fig. \ref{fig:keffnu}. These results indicate that $\kappa_\mathrm{eff}$ decreases monotonically with increasing magnetization with the familiar asymptote at $\hat{\kappa}_\mathrm{eff} = 1/3$, meaning that a fully magnetized hot-spot has thermal conduction that is $1/3$ of the unmagnetized counterpart. In addition, these results show that the effective temperature index $\nu$ in the thermal conduction first decreases sharply from $5/2$ to nearly $3/2$ then increases again with further magnetization.

\begin{figure}[h]
    \centering
    \includegraphics{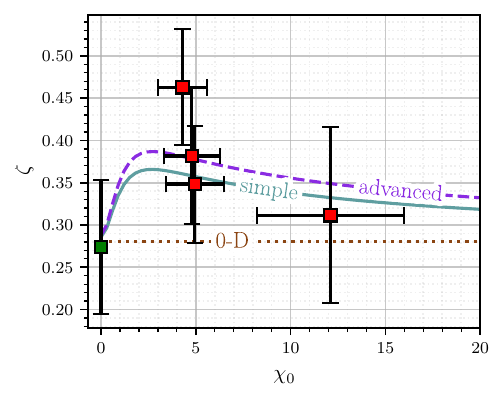}
    \caption{Solutions to the hot-spot temperature profile exponent $\zeta$ using the simple model from Sec. \ref{sec:lindl} and the advanced model from Sec. \ref{sec:betti}. Also shown is the 0-D model from Walsh \textit{et al.} \cite{walshMagnetizedICFImplosions2022} which does not account for the spatial dependence of the temperature and thermal conduction in the hot-spot. Overlaid are the experimentally inferred $\chi_0$ and $\zeta$ for the three experiments \cite{lahmannOneDimensionalPlasma2022}.}
    \label{fig:zeta_withtheory}
\end{figure}

Fig. \ref{fig:zeta_withtheory} shows that both the simple and advanced models have the characteristic increase in $\zeta$ for moderately magnetized hot-spots then decrease with further magnetization. This provides an analytic explanation for the experimental data, which has been plotted with $\chi_0$ inferred from reported values of the central ($r=0$) electron number density and electron temperature \cite{lahmannOneDimensionalPlasma2022}, with volume-average $B$-field assuming perfect flux compression \cite{sioPerformanceScalingApplied2023}.

For convenience, we remark that the temperature profile exponent $\zeta$ determined using the method in this section can be approximately fit by the function
\begin{equation} \label{eq:zetafit}
    \zeta \approx \frac{\chi_0}{8 + 5\chi_0 + 0.8 \chi_0^2} + \frac{2}{7}.
\end{equation}

Put simply, the reason why the temperature profile is most altered for moderate magnetization is that the heat flow is largely suppressed in the hot-spot center, but not near the hot-spot edges, where the Hall parameter is less (we term this phenomenon \textit{differential insulation}). In contrast, highly magnetized hot-spots have large Hall parameters throughout the hot-spot (including the edge) and thus the heat flow is suppressed the same everywhere. Moderately magnetized hot-spots have a peaked temperature profile, meaning that heat is leaking out of the hot-spot edges, but much less heat is leaking out of the center due to magnetized thermal confinement. Notably, this effect is exaggerated in the advanced model, which predicts that the temperature profile peakedness is greater than the simple model (as shown in Fig. \ref{fig:zeta_withtheory}). This might be since mass ablation injects cold shell material into outer region of the hot-spot, decreasing the temperature near the shell and thus increasing the temperature profile peakedness.

\section{Hot-spot model} \label{sec:Tamp}

An important effect of magnetizing implosions is the increase in temperature due to suppression of thermal losses \cite{changFusionYieldEnhancement2011a,moodyIncreasedIonTemperature2022,sioPerformanceScalingApplied2023}. The temperature enhancement due to magnetization in the hot-spot is derived in Walsh \textit{et al.} \cite{walshMagnetizedICFImplosions2022} by making a number of simplifying assumptions. Some of these include (a) the hot-spot mass being dominated by ablation off of the inner surface of the dense fuel layer, (b) hot-spot variables are uniform in the hot-spot (e.g., temperature is homogeneous), (c) pressure and convergence are unaffected by magnetization and (d) ablation is dominant around peak compression. For the NIF gas-filled Symcaps in these experiments, there is no cryogenic dense fuel layer, and so therefore there is no fuel ablation as assumed by (a) and (d). Furthermore, Sections \ref{sec:lindl} and \ref{sec:betti} show that the shape of the temperature profile is important and altered by magnetization, which encourages us to remove assumption (b). Finally, since in a stagnation without ablation into the hot-spot then $n_\mathrm{hs} R_\mathrm{hs}^3 = \mathrm{const.}$ and $p_\mathrm{hs} \sim n_\mathrm{hs} T_\mathrm{hs}$, it is impossible for the temperature to be amplified by magnetization without $R_\mathrm{hs}$ and/or $p_\mathrm{hs}$ changing. These four points show that the assumptions in the Walsh \textit{et al.} derivation are not appropriate for making predictions about NIF gas-filled Symcaps, and motivates our derivation of a new hot-spot model.

The Symcap model we present in this section evolves the essential hot-spot quantities throughout time (temperature, pressure, radius, Hall parameter) during the deceleration phase of the implosion, where the effects of magnetization are greatest. Instead of evolving spatially-resolved variables, we invoke the earlier ansatz of $T(r, t) = T_0(t) (1 - (r/R_\mathrm{hs}(t))^2)^{\zeta(\chi_0)}$. In this framework, the dynamical scalar quantities $T_0$, $R_\mathrm{hs}$, $p_\mathrm{hs}$ and $\chi_0$ completely inform the evolutionary dynamics. This can be readily shown by simplifying the energy equation with this ansatz. After the impulsive deceleration phase and during the continuous deceleration phase, the hot-spot pressure is approximately isobaric \cite{bettiHotspotDynamicsDecelerationphase2001,boseHydrodynamicScalingDecelerationphase2015a}, and so we can write start with the isobaric hot-spot energy equation without alpha-heating.
\begin{equation} \label{eq:energyequation}
    \frac{3}{2} \frac{\partial p}{\partial t} + \frac{5p}{2}\nabla \cdot \vv{u} = \nabla \cdot \left( \kappa_0 T^{5/2} \nabla T \right) - c_b n^2 \sqrt{T}
\end{equation}
Here, $\kappa_0 = (9.81\cdot 10^{7}/\lambda_{ei}) {\mathrm{Mbar \cdot cm^2}}/(\mathrm{s \cdot keV^{7/2}})$ ($\lambda_{ei}$ is the Coulomb logarithm) and $c_b = 5.372\cdot 10^{-36} \mathrm{cm^6 Mbar}/(\mathrm{keV^{1/2} s})$.\cite{hurricanePhysicsPrinciplesInertial2023a} Integrating over the spherical hot-spot volume,  invoking that $\vv{u}|_{r \to R_\mathrm{hs}} = \partial R_\mathrm{hs}/\partial t$ and using the magnetized thermal conduction integral from Eq. \ref{eq:unnormalizedlindl}, we find 

\begin{multline} \label{eq:volumeenergyeqn}
    \frac{3}{2} \frac{4}{3} \pi R^3 \frac{\partial p}{\partial t} + \frac{5}{2} 4 \pi R^2 p \frac{\partial R}{\partial t} = \\
    -4 \pi R^2 \left[ \kappa_0 T^{5/2} \frac{\partial T}{\partial r} \left( \frac{1}{3} + \frac{2}{3} \frac{\kappa_\perp}{\kappa_\parallel} \right) \right]_{r \to R} - c_b \int n^2 \sqrt{T} dV,
\end{multline}
where we have substituted $R$ for $R_\mathrm{hs}$ for brevity. Using our typical ansatz for the temperature profile and assuming the hot-spot is isobaric and follows the ideal gas law (i.e., $n(r) \sim T(r)^{-1}$), the integral in Eq. \ref{eq:volumeenergyeqn} reduces to 
\begin{equation} \label{eq:bremintegral}
    4 c_b \pi  R^3 n_0^2 \sqrt{T_0} \int_0^1 (1 - \hat{r}^2)^{-3\zeta/2} \hat{r}^2 d\hat{r},
\end{equation}
where $n_0$ and $T_0$ are the time-dependent central hot-spot number density and temperature, respectively. By defining the new integral function $\Phi(x)$ as 
\begin{equation} \label{eq:phiintegral}
    \Phi(x) = 3 \int_0^1 \left( 1 - \hat{r}^2 \right)^x \hat{r}^2 d\hat{r} \approx \frac{1}{0.31x^2 + 1.31x + 1}, 
\end{equation}
where the approximation is within 3\% error for $-1 < x < 0.7$ (important since $-0.6 < -\frac{3}{2}\zeta < -0.4$), the resulting equations can be written in a more compact form. A straightforward rearrangement of Eq. \ref{eq:volumeenergyeqn} results in 
\begin{multline} \label{eq:unnormenergyequation}
    \frac{\partial p}{\partial t} = -5 \frac{p}{R} \left( \frac{\partial R}{\partial t} \right)
    - \frac{2 \kappa_0}{R} \left[ T^{5/2} \frac{\partial T}{\partial r} \left( \frac{1}{3} + \frac{2}{3} \frac{\kappa_\perp}{\kappa_\parallel} \right) \right]_{r \to R} \\
    - \frac{2}{3} c_b n_0^2 \sqrt{T_0} \Phi \left( -\frac{3}{2} \zeta \right).
\end{multline}
Closure for the radial evolution of the shell position is given by a piston-like thin shell model \cite{bettiDecelerationPhaseInertial2002,christophersonTheoryIgnitionBurn2020a,zhouMeasurableLawsonCriterion2008a,hurricaneAnalyticAsymmetricpistonModel2020}, where shell inertia is dictated by the hot-spot pressure.
\begin{equation} \label{eq:thinshellmodel}  
    \frac{4}{3} \pi R^2 \delta R \rho_\mathrm{sh} \frac{\partial^2 R}{\partial t^2} = 4 \pi R^2 p
\end{equation}
Here, $\delta R$ is the shell thickness and we can approximate the shell areal density as $(\rho R)_\mathrm{sh} = \rho_\mathrm{sh} \delta R$. In the incompressible shell model, it is assumed that $(\rho R)_\mathrm{sh}$ is constant throughout the deceleration phase, whereas in contrast in the perfectly compressible limit $(\rho R)_\mathrm{sh} \sim R^{-2}$. We conducted 1-D simulations using the \texttt{HYDRA} code \cite{marinakThreedimensionalSimulationsNova1996} of the unmagnetized WarmMag Symcaps and roughly found that a scaling of $(\rho R)_\mathrm{sh} \sim R^{-1.4}$ (between the two extremes) best matched the shell trajectory. This allows us to make computations without invoking the more complicated thick-shell model as in Ref. \onlinecite{bettiDecelerationPhaseInertial2002}.

To solve the coupled differential equations (Eqs. \ref{eq:unnormenergyequation} and \ref{eq:thinshellmodel}), we cast the problem into dimensionless form. Let $\tilde{p} = p/p(0)$, $\tilde{R} = R/R(0)$, $\tilde{t} = -t v(0)/R(0)$, $\tilde{T} = T_0/T_0(0)$, and 
\begin{align}
    \tilde{g} &= \frac{p(0) R(0)}{(\rho R)_\mathrm{sh} v(0)^2} \\
    \tilde{\alpha} &= \frac{2 \kappa_0 T_0(0)^{7/2}}{p(0)R(0)v(0)} \\
    \tilde{\beta} &= \frac{2 c_b n_0(0)^2 \sqrt{T_0(0)} R(0)}{p(0) v(0)},
\end{align}
where, for any hydrodynamic variable $f$, $f(0)$ is the value of $f$ at the onset of the deceleration phase and $f_0$ is the value at the center of the hot-spot. The initial conditions ($\tilde{t} = 0$) are thus $\tilde{p} = 1$, $\tilde{R} = 1$, and $\partial \tilde{R}/\partial \tilde{t} = -1$. Since these Symcaps lack mass ablation into the hot-spot, the hot-spot mass is unchanging with time and with magnetization during the deceleration phase, i.e., $\int n dV = \frac{4}{3} \pi R^3 n_0 \Phi(-\zeta)$ is an invariant. After writing this expression in terms of normalized variables, applying the ideal gas equation of state ($\tilde{p} = \tilde{n} \tilde{T}$), and setting it equal to its value at the onset of the deceleration phase (where $\zeta=2/7$), then we find the closure for the normalized temperature.
\begin{equation} \label{eq:normtemperature}
    \tilde{T} = \tilde{p} \tilde{R}^3 \frac{\Phi \left( -\zeta \right)}{\Phi \left( -2/7 \right)}
\end{equation}
Finally, the normalized versions of Eqs. \ref{eq:unnormenergyequation} and \ref{eq:thinshellmodel} can be written down.

\begin{align}
    \frac{\partial \tilde{p}}{\partial \tilde{t}} &= 
    - \frac{5 \tilde{p}}{\tilde{R}} \frac{\partial \tilde{R}}{\partial \tilde{t}} 
    - \frac{\tilde{\alpha} \tilde{T}^{7/2} \tilde{\Omega} }{\tilde{R}^2}
    - \frac{\tilde{\beta} \tilde{p}^2}{\tilde{T}^{3/2}} \Phi \left( -\frac{3}{2} \zeta \right) \label{eq:hsmodel1}\\
    \frac{\partial^2 \tilde{R}}{\partial \tilde{t}^2} &= \tilde{g} \tilde{p} \tilde{R}^{1.4} \label{eq:hsmodel2} \\
    \tilde{\Omega} &= \left[ \left( \frac{1}{3} + \frac{2}{3} \frac{\kappa_\parallel}{\kappa_\perp} \right) \left( 2 \hat{r} \zeta \left( 1 - \hat{r}^2 \right)^{7\zeta/2 - 1} \right) \right]_{\hat{r} \to 0.95}  \label{eq:hsmodel3}
\end{align}
Written in this form, the variable $\tilde{\Omega}$ is the alteration to the thermal conduction term due to magnetization suppressing the overall heat flow ($\kappa_\perp/\kappa_\parallel$) and magnetization changing the temperature profile shape $\zeta$. The value at $\hat{r} \to 0.95$ is taken since our earlier ansatz for the temperature profile has an unphysical infinite temperature gradient at $\hat{r} \to 1$. This temperature profile exponent $\zeta$ was determined in the previous section (Eq. \ref{eq:zetafit}), and both $\zeta$ and $\kappa_\perp/\kappa_\parallel$ (Eq. \ref{eq:bragheat}) depend on the central Hall parameter $\chi_0(t)$. Assuming uniform magnetic flux compression ($B_0 \sim \tilde{R}^{-2}$), then the central Hall parameter is $\chi_0 = \chi_0(0) \tilde{T}^{5/2} / (\tilde{p} \tilde{R}^2)$, where $\chi_0(0)$ can be calculated using Eq. \ref{eq:hallparameter}.

\section{Results} \label{sec:results}

In Sec. \ref{sec:lindl} and Sec. \ref{sec:betti}, we derived the simple and advanced models for $T(r)$. Then in Sec. \ref{sec:Tamp}, we coupled these results to a new deceleration-phase evolutionary model which gives $T(t, r)$, $p(t)$, $\chi(t, r)$, and $R_\mathrm{hs}(t)$. Now in this section, we process these profiles to predict the temperature and yield amplification due to the applied magnetic field for our collection of models.

In computing the model predictions (initial conditions and dimensionless parameters from Sec. \ref{sec:Tamp}) for the NIF Symcap experiments, the pressure, hot-spot radius, shell areal density, implosion velocity, central temperature and number density are extracted from a 1-D \texttt{HYDRA} \cite{marinakThreedimensionalSimulationsNova1996} simulation of the unmagnetized NIF WarmMag Symcap N210912. In the simulation, the capsule is driven using a frequency-dependent X-ray source tabulated from the 2D \texttt{Lasnex} Hohlraum simulations of N210912 \cite{strozziDesignModelingIndirectly2024}. The 1-D simulation is run through the acceleration phase and the impulsive deceleration phase (where a series of weakening shocks decelerate the imploding shell \cite{bettiHotspotDynamicsDecelerationphase2001}) and the scalar implosion variables listed above are exported. At this point in the implosion, a conventional hot-spot has been formed and so our ansatz for the temperature profile (Eq. \ref{eq:zetaprofile}) is appropriate. It is known that the dominant effect of the magnetic field in these experiments is in the suppression of thermal conduction, which scales with $\chi_0$, and therefore the effect of magnetization is small up until the start of the deceleration phase. Thus, we can assume that the applied magnetic field does not affect the implosion variables at the start of the continuous deceleration phase. For N210912, these variables are $R(0) = 85 \ \mathrm{\mu m}$, $p(0) = 7.8 \ \mathrm{GBar}$, $(\rho R)_\mathrm{sh} (0) = 0.34 \ \mathrm{g/cm^2}$, $v(0) = 180 \ \mathrm{km/s}$, $T_0(0) = 3 \ \mathrm{keV}$ and $\rho_0 (0) = 1.92 \ \mathrm{g/cm^3}$. This set of parameters results in the dimensionless $\tilde{g} = 0.61$, $\tilde{\alpha} = 2.31$ and $\tilde{\beta} = 0.19$. 

\begin{figure}[h]
    \centering
    \includegraphics{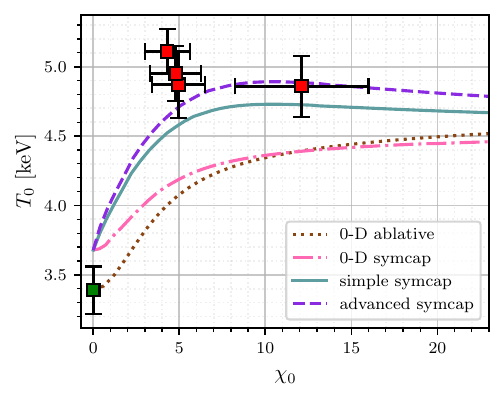}
    \caption{Amplification of the central temperature in magnetized implosions. Lines represent the scalings from four different model predictions (described in Table \ref{tab:models}). The models differ by whether they assume there is a dense (cryogenic) fuel layer to induce ablation, and whether the temperature profile shape (parameterized by $\zeta$) depends on the magnetization $\chi_0$. Experimental data with corresponding error-bars is drawn from the analysis of NIF shots N210912, N210607, and N220110 in Ref. \onlinecite{lahmannOneDimensionalPlasma2022}. The 0-D symcap, simple symcap, and advanced symcap models give actual temperature, whereas 0-D ablative only gives temperature scaling (i.e., $T_0/T_{0,\mathrm{no-B}}$) and is fit to the $B=0$ experiment.}
    \label{fig:centraltemp}
\end{figure}

Figure \ref{fig:centraltemp} shows the results of this Symcap hot-spot model in predicting the central temperature amplification due to hot-spot magnetization. Four models are shown against the data from the NIF experiments. The simplified characteristics of these hot-spot model variations is shown in Table \ref{tab:models}. The 0-D ablative model is from Walsh \textit{et al.},\cite{walshMagnetizedICFImplosions2022} and was derived by assuming that ablation off of the fuel layer into the hot-spot is dominant, whereas the other models use Eqs. \ref{eq:hsmodel1}-\ref{eq:hsmodel3} and are therefore appropriate for modeling Symcaps. Both of the first two models assume that the temperature profile is unaffected by the magnetization (i.e., $\zeta = \mathrm{const.}$). The last two models both model the change in temperature profile shape due to magnetization, but the ``simple'' symcap model is based on the results from Sec. \ref{sec:lindl}, whereas the ``advanced'' symcap model is based on Sec. \ref{sec:betti} and is anticipated to be more accurate. It is worth noting that the value of $\kappa_\mathrm{eff}$ we use for the 0-D ablative \cite{walshMagnetizedICFImplosions2022} model is calculated using the Hall parameter near the edge of the hot-spot ($r/R_\mathrm{hs} = 0.95$) since this is the location of ablation surface that dictates the hot-spot power balance.

\begin{table}
\caption{\label{tab:models}Summary description of hot-spot models used to predict temperature amplification (Fig. \ref{fig:centraltemp}) and yield amplification (Fig. \ref{fig:rawyield}).}
\begin{ruledtabular}
\begin{tabular}{cccc}
    Model name & Fuel-layer & Model equation & Formula for $\zeta$\\
    & ablation & & \\
    \hline
    0-D ablative & Yes, dominant & Ref. \onlinecite{walshMagnetizedICFImplosions2022}, Eq. 9 & $\displaystyle \zeta = 2/7$ \\
    0-D symcap & No & Eqs. \ref{eq:hsmodel1}-\ref{eq:hsmodel3} & $\displaystyle \zeta = 2/7$ \\
    simple & No & Eqs. \ref{eq:hsmodel1}-\ref{eq:hsmodel3} & $\displaystyle \frac{2}{7} + \frac{\chi_0}{12 + 5 \chi_0 + 1.3 \chi_0^2}$ \\
    advanced & No & Eqs. \ref{eq:hsmodel1}-\ref{eq:hsmodel3} & $\displaystyle \frac{2}{7} + \frac{\chi_0}{8 + 5 \chi_0 + 0.8 \chi_0^2}$ \\
\end{tabular}
\end{ruledtabular}
\end{table}

The results in Fig. \ref{fig:centraltemp} suggest that including the change in temperature profile due to magnetization is a key aspect to accurately predicting the central temperature amplification. Both of the 0-D models underpredict the temperature amplification for the magnetized experiments. Furthermore, the 0-D ablative model (from Walsh \textit{et al.} \cite{walshMagnetizedICFImplosions2022}) only provides a scaling ratio (e.g., $T/T_\mathrm{no-B}$) and not the actual temperature or yield. To predict the temperature with the 0-D ablative model one needs to specify $T_\mathrm{no-B}$---this is usually taken from an experiment and is sensitive to experimental uncertainty.

\begin{figure}[h]
    \centering
    \includegraphics{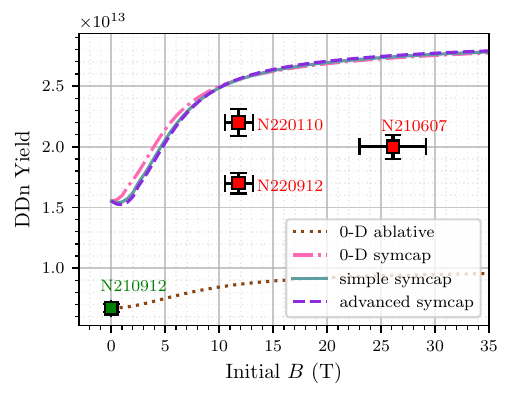}
    \caption{Fusion yield predictions by the four models described in Table \ref{tab:models}. The legend is equivalent to Fig. \ref{fig:centraltemp}. Experimental measurements for shots N210912, N220110, N210607, and N220912 are shown with errorbars and shot number labels. The 0-D symcap, simple symcap, and advanced symcap models give actual yield, whereas 0-D ablative only gives yield scaling (i.e., $Y/Y_\mathrm{no-B}$) and is fit to the $B=0$ experiment.}
    \label{fig:rawyield}
\end{figure}

In Fig. \ref{fig:rawyield}, the experimental neutron yields for the WarmMag campaign is compared to the yield predicted by the 0-D ablative, 0-D symcap, simple symcap and advanced symcap models. There are four curious features shown by this scaling plot, each of which will be discussed at length throughout the rest of this section.
\begin{enumerate}
    \item All three of the Symcap models predict similar yields, despite having different central temperature amplification; this is a coincidence due to the parameter space of these experiments.
    \item The Symcap models overestimate the measured yield for all experiments; this is likely due to degradative mix into the hot-spot.
    \item Experiment N210912 ($B=0$) is especially fit poorly by the Symcap models; this implosion was more oblate than the others and received less laser drive energy.
    \item Experiment N220110 ($B=12$ T) has higher yield than the others and does not follow the yield trend; this shot had a capsule leak and near-vacuum hohlraum fill and is not readily comparable to the other experiments.
\end{enumerate}

Let us start by addressing the first point, that even though the three Symcap models predict different central temperature amplification, they all predict similar yields. We will show that this is a coincidence, where the 1-D profile shape effect increases the central temperature but decreases the effective burn volume by a similar amount, so that the yield remains the same. To illustrate this, let us write down the generic formula for the fusion yield, parameterizing the fusion cross-section $\langle \sigma v \rangle$ in terms of an approximate power law \cite{boschImprovedFormulasFusion1992,richardson2019NRLPlasma2019}: $\langle \sigma v \rangle_{DD} \approx 6.7 \cdot 10^{-22} (T/\ \mathrm{keV})^{3.3} \ \mathrm{cm^3/s}$ (valid for $3 \ \mathrm{keV} < T < 6 \ \mathrm{keV}$).
\begin{equation} \label{eq:yieldgeneric}
    Y = \iint \frac{n^2}{2} \langle \sigma v \rangle_{DD} dV dt,
\end{equation}
Invoking particle number conservation in the hot-spot, the central number density is 
\begin{equation} \label{eq:numberconservation}
    n_0(t) = n_0(0) \left( \frac{R(0)}{R(t)} \right)^3 \left( \frac{\Phi(-2/7)}{\Phi(-\zeta(t))} \right).
\end{equation}
The scaling for the fusion yield due to shape effects and central temperature amplification is found by: writing Eq. \ref{eq:yieldgeneric} in terms of normalized variables (from Sec. \ref{sec:Tamp}), using the power-law fit for $\langle \sigma v \rangle_{DD}$, inserting the central number density from Eq. \ref{eq:numberconservation}, assuming a constant burn-width, then taking the ratio compared to the unmagnetized case.
\begin{equation} \label{eq:yieldscaling}
    \frac{Y}{Y_{B=0}} \approx 3.61 \left( \frac{R_{B=0}}{R} \right)^3 \frac{\Phi(1.3 \zeta)}{(\Phi(-\zeta))^2} \left( \frac{T_0}{T_{0,B=0}} \right)^{3.3}
\end{equation}
Shown by this equation, the scaling of the yield with magnetization depends on (1) the difference in hot-spot volume (assumed to be small), (2) two profile shape factors, where the numerator comes from the integral of \eqref{eq:yieldgeneric} over the hot-spot volume and the denominator comes from number conservation (Eq. \ref{eq:numberconservation}), and (3) the power-law amplification of the central temperature. To get a sense of what this shape factor (term 2) looks like, we note that it can be approximated by
\begin{equation}
    3.61 \frac{\Phi(1.3 \zeta)}{(\Phi(-\zeta))^2} \approx 1 - 3.8 \left( \zeta - \frac{2}{7} \right),
\end{equation}
showing that the shape factor decreases when the temperature profile gets more peaked (i.e., higher $\zeta$). The combined effect is this: when profile shape effects are included in the models, the central temperature goes up (see Fig. \ref{fig:centraltemp}), but including shape effects makes the yield go down. It is our claim that these terms nearly balancing one another is coincidental, and that implosions with higher temperatures or different deceleration-phase conditions will not have unchanging yields when the profile shape is included.

Demonstrated in Fig. \ref{fig:rawyield} is that all three of the Symcap hot-spot models overestimate the experimental yield. Prior \texttt{Lasnex} simulations of these experiments \cite{strozziDesignModelingIndirectly2024} also overpredict the yield by a factor of 2-3x, suggesting the discrepancy is not just a feature of our semi-analytic model. Ongoing research aims to identify the physics source of this discrepancy, although a leading candidate is hydrodynamic mixing between the high-Z shell and the hot-spot during the deceleration phase of the implosion \cite{welser-sherrillApplicationFalllineMix2008a,smalyukMeasurementsAblatorGasAtomic2014,zylstraHotspotMixLargescale2020}. Since this discrepancy between simulations and experiments occurs in all cases (magnetized and unmagnetized), the cause is likely systematic, i.e., unrelated to the magnetic field. Assuming that the high-Z mix is annular and the enhancement in Bremsstrahlung emission due to the mix effectively nullifies the yield in the mixed region by decreasing the temperature, then we can estimate the yield by replacing $\Phi(x)$ in the numerator of Eq. \ref{eq:yieldscaling} with a ``mixed'' profile integration:
\begin{equation} \label{eq:psimix}
    \Psi(x, \hat{w}) = 3 \int_0^{1-\hat{w}} (1-\hat{r}^2)^x \hat{r}^2 d\hat{r}.
\end{equation}
In this equation, $\hat{w}$ is the time-dependent mix width normalized to the hot-spot radius. Then the degradation of yield due to mixing is given by 
\begin{equation} \label{eq:mixedyield}
    \frac{Y}{Y_\mathrm{no-mix}} \approx \left( \frac{\Psi(1.3\zeta, \hat{w})}{\Psi(1.3\zeta, 0)} \right).
\end{equation}
Hydrodynamic mixing in ICF is generally seeded by the Rayleigh-Taylor or Richtmyer-Meshkov instabilities, which in their non-linear regimes the mixing width scales like $\hat{w} \sim t^2$, according to fall-line \cite{welser-sherrillApplicationFalllineMix2008a} and buoyancy-drag analyses \cite{dimonteKLTurbulenceModel2006a}. However, for simplicity we will assume that $\hat{w}$ is constant over the burn width (i.e., the integration time of Eq. \ref{eq:yieldgeneric}). Given that the discrepancy between the models and the experimental data for shot N210607 (26.1 T) is approximately $Y/Y_\mathrm{no-mix} \approx 0.74$, then solving Eq. \ref{eq:mixedyield} implies that $\hat{w} \approx 0.15$. This suggests that the outer 15\% of the hot-spot by radius (or 39\% by volume) would be polluted by mix to completely explain the degradation to the yield. Other leading candidates for the yield degradation include low-mode hot-spot shape distortions (mode-2 and mode-4), mix injected by the fill-tube, or a bulk hot-spot flow (mode-1). For fill-tube mix, simple assessment of Eq. \ref{eq:psimix} for $x = 1.3 \cdot 2/7$ suggests that the majority of fusion reactions occur for $\hat{r} > 0.8$, and that localized mix in the center of the hot-spot may play a small role. However, for the fill-tube jet to penetrate into the center, it necessarily pollutes mix into some fraction of the outer radii of the hot-spot.

\begin{figure}[h]
    \centering
    \includegraphics{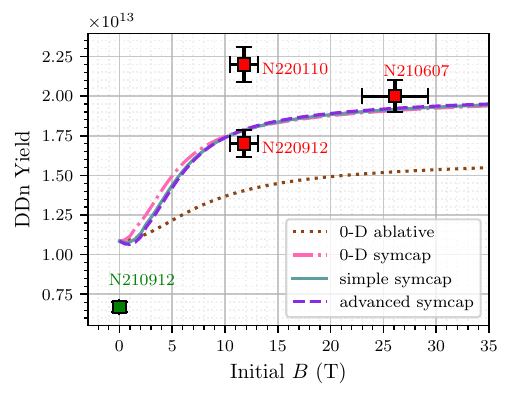}
    \caption{Model predictions of the fusion yield after rescaling the predictions of Fig. \ref{fig:rawyield} to fit the magnetized experiments (all symcap models multiplied by 0.75), motivated by systematic yield degradation. The 0-D ablative model \cite{walshMagnetizedICFImplosions2022} is renormalized to align with the new unmagnetized prediction. Experimental measurements for shots N210912, N220110, N210607, and N220912 are shown with errorbars and shot number labels.}
    \label{fig:yieldscaling}
\end{figure}

In Strozzi \textit{et al.} \cite{strozziDesignModelingIndirectly2024}, symmetrized simulations of shot N210607 with and without a magnetic field suggests that the magnetic field alone gives a yield enhancement of 1.65x, and that low-mode shape distortions and decrease in laser drive degraded N210912 by $\sim38\%$ ($(1/1.62)$x) compared to the analogous N210607 experiment. Therefore, instead of renormalizing all of the models in Fig. \ref{fig:rawyield} to fit through the $B=0$ experiment (which had numerous degradations), we fit it to $1.62 \times Y_\mathrm{N210912}$ to remove the systematic degradation (to all experiments) without accruing additional error from the additional degradations in the $B=0$ experiment. Some candidates for the additional degradations affecting N210912 compared to the other experiments include:
\begin{itemize}
    \item Two dropped laser quads, which decreases total laser energy by 5\% and increases implosion asymmetry.
    \item Decreased implosion symmetry compared to the magnetized case. The unmagnetized shot N210912 was more oblate ($P_2/P_0 = -16.6\%$) than the 26 T shot N210607 ($P_2/P_0 = 5.7\%$) \cite{moodyIncreasedIonTemperature2022}. Magnetization changes the symmetry by altering the hohlraum physics and through anisotropic thermal conduction in the hot-spot \cite{strozziDesignModelingIndirectly2024}.
    \item Hydrodynamic mixing may affect magnetized implosions less than unmagnetized ones. Magnetic tension can oppose the growth of the Rayleigh-Taylor and Richtmyer-Meshkov instabilities, which would suppress mixing \cite{walshMagnetizedAblativeRayleighTaylor2022,walshResistiveDiffusionMagnetized2024,perkinsPotentialImposedMagnetic2017}.
\end{itemize} 

Finally, it is worth addressing why shot N220110 (initial B of 12 T) in Fig. \ref{fig:yieldscaling} is not predicted well by the models. This shot had a capsule leak and no neopentane hohlraum gas fill \cite{sioPerformanceScalingApplied2023}. One possibility for the higher yield in this experiment is that the lower hohlraum gas fill allowed more energy coupling to the capsule \cite{hallRelationshipGasFill2017}, however, \texttt{Lasnex} simulations of this shot were not conducted since the near-vacuum hohlraum gas fill complicates modeling. Instead, shot N220912 which also had a 12 T field is considered to be more comparable to the other experiments for scaling purposes.

\section{Conclusion} \label{sec:conclusion}

In this work, we present hot-spot models relevant to the NIF Symcaps indicating that the radial temperature profile is altered for implosions with an axial magnetic field, and show that this effect is important to accurately modeling the hot-spot temperature and fusion yield. We first determined the shape of the temperature profile of a magnetized hot-spot, using a \textit{simple} (Sec. \ref{sec:lindl}) illustrative steady-state power-balance model and an \textit{advanced} (Sec. \ref{sec:betti}) model which includes convergent, non-steady-state dynamics. Both of these models indicate that moderately magnetized (i.e., with central electron Hall parameter $2 < \chi_0 < 10$) implosions have a more peaked radial temperature profile compared to unmagnetized and highly magnetized implosions, which matches the results from NIF experiments (Fig. \ref{fig:zeta_withtheory}). This effect occurs since, in moderately magnetized hot-spots, heat flux is more suppressed within the core of the hot-spot compared to the hot-spot edges, insulating only part of the hot-spot and creating a peaked temperature profile. In contrast, the strongly magnetized hot-spots are uniformly insulated; this lack of differential insulation means that strongly magnetized hot-spots have similar profile shapes to unmagnetized hot-spots, but with higher overall temperatures. In Sec. \ref{sec:Tamp}, we derived a dynamic hot-spot model that uses our results for the temperature profile shape exponent $\zeta(\chi_0)$, and the four hot-spot scalar parameters ($R$, $T_0$, $p$, $\chi_0$) were evolved throughout the deceleration phase by self-consistently including the profile shape effects on the thermal conduction and Bremsstrahlung emission. 

We then show that these models match the measurements of temperature profiles and central temperature amplification in NIF experiments (Fig. \ref{fig:centraltemp}) \cite{lahmannOneDimensionalPlasma2022}. Discrepancies between the predicted and measured yield (shown in Fig. \ref{fig:rawyield}) are discussed, and it is suggested that systemic degradations such as ablator mixing into the hot-spot degrade all experimental yields by $\approx 25\%$ (where Fig. \ref{fig:yieldscaling} shows the rescaled model predictions), but the $B=0$ reference experiment had enhanced degradation due to decreased laser drive, increased hot-spot oblateness, and possibly lacked the magnetized suppression of instability growth. Ongoing work aims to close this discrepancy in the yield modeling.

The hot-spot model presented in this work motivates further research into understanding magnetized implosions. The change in the temperature profile shape may have ramifications on the alpha-heating characteristics in future cryogenic implosions or ablator mixing rate in all magnetized implosions. The model presented in this work will also be used to interpret diagnostic measurements for future magnetized experiments at the NIF. Ongoing work seeks to assess the behavior of this model in 2D, where the anisotropy in the thermal conduction and alpha heating will induce an anisotropy in the temperature profiles.

\begin{acknowledgments}
    This material is supported by Department of Energy National Nuclear Security Administration Stewardship Science Graduate Fellowship program under award number DE-NA0003960, and through the University of Rochester Laboratory for Laser Energetics subaward number DENA0003856: SUB00000056/GR530167/AWD00002510. Partially supported by LLNL LDRD 23-ERD-025. 
\end{acknowledgments}

\section*{Author Contribution}
R. C. Spiers: Conceptualization (lead), Investigation (lead), Methodology (lead), Writing/Original Draft Preparation (lead), Writing/Review \& Editing (equal)
A. Bose: Conceptualization (equal), Investigation (supporting), Methodology (equal), Supervision (lead), Writing/Review \& Editing (lead)
C. A. Frank: Investigation (supporting), Methodology (supporting), Writing/Review \& Editing (supporting)
B. Lahmann: Data Curation (lead), Conceptualization (supporting), Investigation (supporting), Writing/Review \& Editing (supporting)
J. D. Moody: Investigation (supporting), Validation (equal), Writing/Review \& Editing (equal)
H. Sio: Data curation (equal), Investigation (supporting), Writing/Review \& Editing (supporting)
D. J. Strozzi: Investigation (supporting), Validation (equal), Writing/Review \& Editing (equal)

\bibliography{magtemp}

\end{document}